\documentclass{PoS}

\newcommand{\be}{\begin{equation}}
\newcommand{\ee}{\end{equation}}
\newcommand{\bea}{\begin{eqnarray}}
\newcommand{\eea}{\end{eqnarray}}

\newcommand{\Pg}{\hat{{\rm P}}_{\rm G}}
\newcommand{\Tr}{\,\hbox{\rm Tr}}

\title{Glueball masses with exponentially improved statistical
   precision}

\ShortTitle{Glueball masses with exponentially improved statistical
   precision}

\author{\speaker{Michele Della Morte}\\
       Institut f\"ur Kernphysik and Helmholtz Institut, University of Mainz,\\
       Johann-Joachim-Becher Weg 45, D-55099 Mainz, Germany\\
       E-mail: \email{morte@kph.uni-mainz.de}}

\author{Leonardo Giusti\\
       Dipartimento di Fisica, Universit\'a di Milano Bicocca, \\
       Piazza della Scienza 3, I-20126 Milano, Italy\\
       E-mail: \email{leonardo.giusti@unimib.it}}

\abstract{
\vspace{-11.5cm}
We briefly review the computational strategy we have recently
 introduced for computing glueball masses and matrix elements, which 
 achieves an exponential reduction of statistical errors compared to 
standard  techniques. The global symmetries of the theory play a crucial 
role in the  approach. We show how our previous work on parity can be
 generalized to other  symmetries.
In particular we discuss how to extract the mass of the $0^{++}$, 
$2^{++}$ and
$0^{-+}$ lightest glueballs avoiding the exponential degradation of the 
signal to noise ratio. We present new numerical results and update the
 published ones.
\vspace{-16.cm}
\begin{flushright}
MKPH-H-10-27\\
\end{flushright}
}

\FullConference{The XXVIII International Symposium on Lattice Field Theory, Lattice2010\\
		June 14-19, 2010\\
		Villasimius, Italy}

\begin{document}
\section{Introduction}
The self-coupling of gluons in quantum chromodynamics  suggests the
 existence of glueballs, bound states of mainly gluons. A clear 
experimental evidence for their existence remains however elusive. On the
theoretical side predictions for the glueball masses are very difficult
to obtain. They require a detailed knowledge of the QCD vacuum,
which cannot be obtained by standard perturbative techniques. The most 
reliable approach is provided by numerical simulations of the theory on
a space-time lattice.
In general, the mass of the lightest asymptotic
state with a given set of quantum numbers can be extracted from
the Euclidean time dependence of a suitable two-point correlation
function computed on the lattice via Monte-Carlo simulations.
The contribution of the lightest state can be disentangled from those of
other states by inserting the source fields at large-enough time 
distances. The associated statistical error can be estimated
from the spectral properties of the 
theory~\cite{Parisi:1983ae,Lepage:1989hd}.
Very often the latter grows exponentially with the time separation,
and in practice it is not possible to find a window where statistical 
and systematic errors are both under control.
This is the major limiting factor in numerical
computations of the glueball masses
in the Yang--Mills theory.
In the following we will review the ``symmetry constrained'' approach we 
have introduced in~\cite{DellaMorte:2008jd} and present numerical 
applications, showing that
it indeed solves the exponential noise to signal problem and it allows
for a complete computation, in principle, of the glueball spectrum.

\section{Symmetry Constrained Monte Carlo}

The exponential noise to signal problem affects the standard procedure
since for any given gauge configuration all asymptotic states of the
theory are allowed to propagate in the time direction, regardless of the
quantum numbers of the source fields. The correct quantum numbers are 
recovered in the gauge average only and as a result of large 
cancellations.
Inspired by the transfer matrix formalism, we have designed a 
multilevel  algorithm in which the propagation in time of states with 
a given set of quantum numbers only is permitted.
The exponential problem is removed in the same way as originally proposed
in~\cite{Luscher:2001up} for  the correlator of Polyakov 
loops. For the sake of simplicity we will here discuss again the case of
the Parity-Symmetry Constrained algorithm and generalize it to all 
lattice symmetries in the following. Hierarchical (or multilevel) 
integration schemes are usually designed for  specific (classes of) 
observables. In our case we want to compute
\be
Z^{-}(T) = \Tr\left\{\hat{\rm T}^T\left({{1+\hat\wp}\over{2}}\right)
\Pg \right\} \;,
\ee
i.e. the contribution of the parity odd states to the partition function.
In the above equation $\hat{\rm T}$ is the Transfer matrix, and T is the 
temporal extent of the lattice, $\Pg$
is the projector on gauge invariant states and $\hat\wp$ is the parity 
transformation operator. In order to build a hierarchical integration scheme
starting form a local update procedure one needs to bring the observable in a 
factorized form. 
We consider an obvious factorized form of $Z$ and generalize it to $Z^{-}$. 
If we introduce thick time-slices of temporal extent $d$ then in the 
configuration basis $Z$ can be written as
\be
Z(T)=\int \prod_{i=0}^{T/d-1} {\bf D}_3[V_{id}] {\rm T}^d[V_{(i+1)d}, V_{id}]\;,
\label{Zfact}
\ee
with
\be
{\rm T}^d[V_{x_0+d}, V_{x_0}]=\langle V_{x_0+d} | \hat{\rm T}^d \Pg |
 V_{x_0} \rangle \;.
\ee
By introducing 
\be
({\rm T}^-)^d[V_{x_0+d}, V_{x_0}]={{1}\over{2}} \left\{ {\rm T}^d[V_{x_0+d}, V_{x_0}]
- {\rm T}^d[V_{x_0+d}, V^\wp_{x_0}] \right\}
\label{todd}
\ee
where the superscrit $\wp$ means that the state has been parity transformed, 
equation~\ref{Zfact} gets immediately generalized to $Z^-(T)$. It is easy to see that
the modified transfer matrix elements $({\rm T}^-)^d[V_{x_0+d}, V_{x_0}]$ vanish
whenever either one of the states $V_{x_0}$ and $V_{x_0+d}$ is invariant under
parity, in other words only parity odd states are transfered. The details of the
numerical implementation of the resulting multilevel algorithm are discussed 
in~\cite{DellaMorte:2008jd}. 
Aiming at the ratio $Z^-/Z$, the basic quantity  to be
computed for each  $sub$-lattice of time extent $d$ with
Dirichlet boundary conditions  is the ratio of partition functions 
\[
{{{\rm T}^d\Big[V_{x_0+d}^\wp,V_{x_0}\Big]}\over{{\rm T}^d\Big[V_{x_0+d},V_{x_0}\Big]}}\,.
\]
The product over the thick time-slices of a simple linear combination of those ratios
is then integrated numerically over the boundary configurations
$V_{x_0=id}$ generated with the usual Boltzmann weight.
An exponential error reduction is achieved because for large 
enough values of $d$, say $d$ of the order of the inverse critical temperature,
each factor is of the right size $e^{-E^-d}$, with $E^-$ the energy of
the lightest parity odd state, on any gauge configuration and its fluctuations
 are reduced to the same level.
\subsection{Results from the parity constrained Monte Carlo}
\begin{figure}[htb]
\begin{center}
\includegraphics[angle=-90,width=10.5cm]{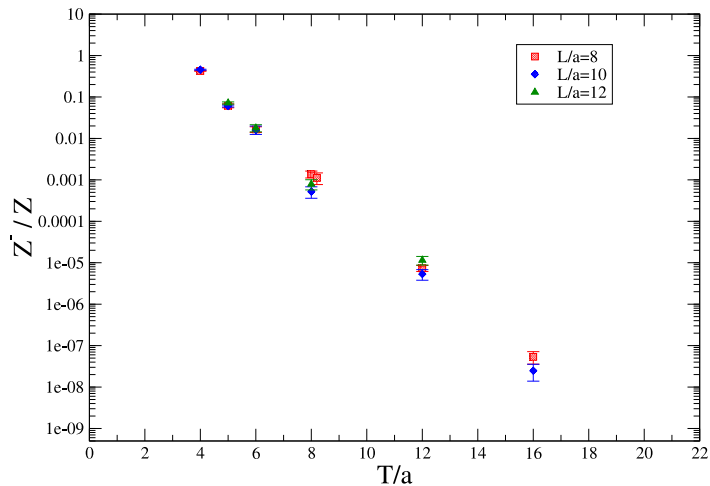}
\caption{The ratio $Z^-/Z$ as a function of $T/a$ for different lattice sizes. 
The lattice spacing is always $a\simeq 0.17$ fm.}
\label{summaryP}
\end{center}
\end{figure}

We have tested this strategy in the Wilson regularization of the SU(3) Yang--Mills 
theory~\cite{Wilson:1977nj}  by determining the relative
contribution to the partition function of the parity-odd states on lattices with a spacing
of roughly $0.17$~fm, spatial volumes up to $8.5\, \mbox{fm}^3$, and time extents up
to $2.7\, \mbox{fm}$~\cite{DellaMorte:2008jd,DellaMorte:2009rf}. 
We show in Fig.~\ref{summaryP} the results for the ratio $Z^-/Z$ as a 
function of $T/a$ for lattices of different size, namely $L/a=8,10$ and $12$.
After tuning the parameters of the algorithm we found 50 to 100 measurements to be enough to 
reach the accuracy shown. We have used a two level algorithm for $T/a\leq 10$ and a higher, 
three level, scheme for $T/a \geq 12$.
These results demonstrate that the algorithm behaves as expected, and that the multi-level 
integration scheme indeed achieves an exponential reduction of the numerical effort with 
respect to the standard procedure. We find quite remarkable  that we could follow an exponential
decay over almost 7 orders of magnitude and for time separations of about 3 fm.
In addition the agreement among the results for different values of $L/a$ shows that
at these volumes torelon contributions, if at all present, are negligible. 
\section{Generalization to the complete set of lattice symmetries} 
Without fixing a complete set of quantum numbers particle identification
is very difficult, even if $T$ is large enough for
the partition function restricted to a given sector to be dominated by a single state.
For example, in the parity odd sector a superposition of states as
\[
{{1}\over{\sqrt{2}}}\left( | 0^{++},\vec{p} \rangle - | 0^{++},-\vec{p} \rangle
\right) \;, \quad |\vec{p}|=2\pi/L
\]
is contributing and its energy  may be smaller than the mass of the lowest parity odd glueball.
Such a contribution could of course be removed by projecting on zero spatial momentum.
The example shows that for a precision study of the glueball spectrum it is
necessary to fix several quantum numbers. The symmetry groups of the lattice SU(3)
Yang-Mills theory are
\begin{itemize}
\item Charge and Parity conjugation. These are Abelian groups composed by 2 elements each.
\item Spatial translations. This is again an Abelian group with $(L/a)^3$ elements.
\item Central Charge conjugation. This is a symmetry group of the pure gauge theory in a finite
volume with periodic boundary conditions~\cite{torelon}. The group is Abelian and is made of
27 elements.
\item The Octahedral group of lattice spatial rotations. It is a non-Abelian group composed
by 24 transformations. 
\end{itemize}
In general elements from different groups do not commute and  therefore only a subset of the 
associated quantum numbers can be simultaneously fixed.
In the following we will show how projectors on irreducible representations of each of the
above groups can be introduced in the formalism presented in the previous section~\cite{paperIII}. 

Let us consider one symmetry group of order $g$. The phase space of the 
theory can be factorized into regular representations constructed by
starting from a state $|V \rangle$ and by applying on it all group 
transformations to obtain the ``vectors'' 
$|V^{\Gamma_i} \rangle = \hat{\Gamma}_i| V \rangle$, $\;i=1\dots g$.
Group theory then tells us how to construct the projector 
$\hat{P}^{(\mu)}$ on 
the irreducible representation $\mu$ of the group, namely
\be
\hat{P}^{(\mu)}= {{n_\mu}\over{g}} \sum_{i=1}^g \chi^{(\mu)*}_i
\hat{\Gamma}_i \;,
\ee
where $n_\mu$ is the dimension of the irreducible representation,
and $\chi^{(\mu)}_i$ is the character of the $i$th group element
in that representation. The contribution $Z^{(\mu)}$ to the partition 
function coming from the states transforming in the irreducible 
representation $\mu$
\be
Z^{(\mu)}(T) = \Tr\left\{\hat{\rm T}^T \hat{P}^{(\mu)}
\Pg \right\} \;,
\ee
can be re-written, in fully analogy to what has been done for the case
of the partition function restricted to the parity odd states, as 
(cf eqs.~\ref{Zfact}-\ref{todd})
\be
Z^{(\mu)}(T)=\int \prod_{i=0}^{T/d-1} {\bf D}_3[V_{id}] \left({\rm T}^{(\mu)}\right)^{\!\!d\!}[V_{(i+1)d}, V_{id}]\;,
\label{Zfactmu}
\ee
where 
\be
\left({\rm T}^{(\mu)}\right)^{\!\!d\!}[V_{x_0+d}, V_{x_0}]={{n_\mu}\over{g}} \sum_{i=1}^g
\chi^{(\mu)*}_i {\rm T}^d[V_{x_0+d}, V^{\Gamma_i}_{x_0}]\;, 
\ee
which extends eq.~\ref{todd}  and reproduces it in the case of the 
non-singlet representation of the parity group. The basic quantities to
be computed in the multilevel algorithm are now the ratios
\[
 {{{\rm T}^d\Big[V_{x_0+d}^{\Gamma_i},V_{x_0}\Big]}\over{{\rm T}^d\Big[V_{x_0+d},V_{x_0}\Big]}}\;, \quad i=1 \dots g \;,
\]
and the product over the thick time-slices of proper linear combinations
of such quantities is then again integrated over the boundary 
configurations.
When several quantum numbers are fixed the number of ratios to be 
computed grows as the product of the dimensions of the associated
groups. The numerical cost of the algorithm however does not increase 
accordingly, basically because the accuracy on each single ratio can be
relaxed as their number becomes larger~\cite{paperIII}.

\section{Numerical Results}

\begin{figure}[htb]
\begin{center}
\includegraphics[angle=-90,width=10.5cm]{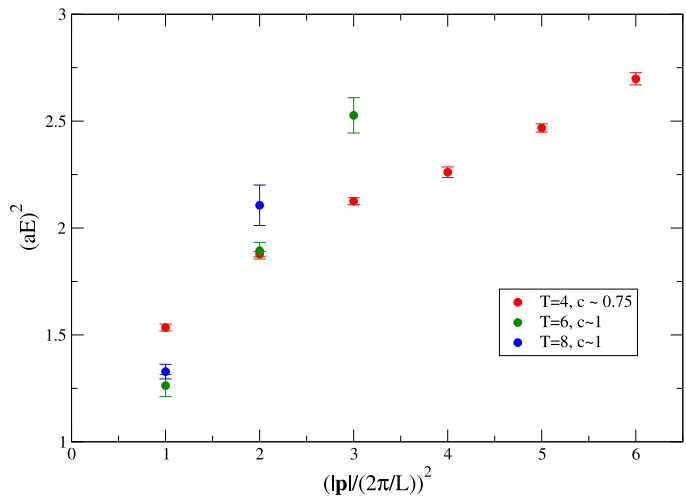}
\caption{Effective energy from the ratio $Z^{\vec{p},C=+}/Z$ (see text) versus
$|\vec{p}|^2$ for different $T/a$.}
\label{disp}
\end{center}
\end{figure}

As an explicit example we consider  the projector on a given spatial
momentum $\vec{p}$. That will be used in most of the cases 
to project to zero momentum except for the computation of the mass
of the lightest $0^{++}$ glueball, where by projecting on a finite 
momentum we will get rid of the otherwise dominating vacuum contribution.
The relative contribution of states with momentum $\vec{p}$ in the 
system with Dirichlet boundary conditions is given by 
($\hat{P}(\vec{x})$ representing translations by $\vec{x}$)
\be
{{(T^{\vec{p}})^d\Big[V_{x_0+d},V_{x_0}\Big]}\over{T^d\Big[V_{x_0+d},V_{x_0}\Big]}}={{1}\over{\sqrt{(L/a)^3}}} 
\sum_{\vec{x}} e^{-i\vec{p}\cdot\vec{x}}
{{T^d\Big[V^{P(\vec{x})}_{x_0+d},V_{x_0}\Big]}\over{T^d\Big[V_{x_0+d},V_{x_0}\Big]}} \;.
\ee
We have implemented the above projection together with the projection on 
the charge conjugation even sector for some of the $L/a=8$ runs 
discussed in  Section~2. This serves as a test of part of the strategy we plan 
to follow to eventually compute the mass of the lowest $0^{++}$ glueball
 state. Results are shown in Fig.~\ref{disp}. We plot the square of an 
effective energy, defined as $-(\ln(Z^{\vec{p},C=+}/Z))/T$ averaged
over the number of momenta $\vec{p}$ with the same $|\vec{p}|^2$, versus 
$|\vec{p}|^2$ for different values of $T/a$.
As $T$ becomes large enough for $Z^{\vec{p},C=+}/Z$ to be dominated by
a single one-particle state the plot should reproduce a linear dispersion 
relation and the slope should approach the speed of light value $c^2=1$.
We see that such a situation is realized for $T/a=6$ and 
larger values of $T$ produce consistent results tough with larger errors.
By extrapolating the $T/a=6$ data to zero momentum we could estimate the
mass of the corresponding state with a few percent accuracy. 
To claim that this state is the $0^{++}$ glueball we however 
need additional
computations in order to exclude parity odd- and higher 
spin-glueballs~\cite{paperIII}.
\section{Conclusions and outlook}
The exponential growth of the noise to signal ratio is one of the main 
obstructions to precise computations in lattice QCD of observables defined 
beyond the mesonic sector.
In the pure gauge theory this problem can be cured by moving away from 
importance sampling and the standard approach relying on the computation
of two-point functions. We have proposed an algorithm where projectors
can be introduced which allow
the propagation in time of states with selected quantum numbers only
on each gauge configuration. The approach has been tested numerically in the 
four-dimensional SU(3) Yang-Mills theory by computing the relative contribution
of parity odd states to the partition function. In this way we could
follow an exponential decay over 7 orders of magnitude and up to 
time-separations of $3$ fm and verify that the algorithm scales as a power
of the time separation for a fixed precision on the rate of the exponential
decay.

We are now performing a glueball spectroscopy study in this setup.
To this end we have extended the approach to include projectors on all lattice
quantum numbers. As a feasibility study of a strategy for the computation
of the mass of the lightest $0^{++}$ glueball we have tested the projector
on a finite spatial momentum. The results are encouraging, however such a 
projector is too expensive to be treated exactly and we are currently
exploring the efficiency of stochastic implementations~\cite{paperIII}.

Finally, compared to other approaches, the one we are following offers the
 possibility to compute in addition
 the multiplicity of a state, and even more interestingly, to numerically
prove the existence of a mass gap in the pure gauge theory~\cite{paperIII}.

\vspace{0.3cm}
\noindent {\bf Acknowledgements.} The simulations were performed on PC 
clusters at CERN,
at CILEA, at the Swiss National Supercomputing Centre (CSCS) and
at the J\"ulich Supercomputing Centre (JSC). We thankfully acknowledge
the computer resources and technical support provided by all these
institutions and their technical staff.

\end{document}